\documentstyle[psfig]{l-aa}

\begin{document}
 
\thesaurus{13
            (11.01.2;  
             11.17.3;  
             13.18.1)} 

\title{Deceleration of Relativistic Radio Components and the morphologies of
Gigahertz Peaked Spectrum Sources.}
 
\author{I.A.G. Snellen \inst{1}, \inst{2},
        R.T. Schilizzi \inst{1}, \inst{3},
         A.G. de Bruyn \inst{4}, \inst{5},
         G.K. Miley \inst{1}} 
 
\institute{ \inst{1} Leiden Observatory, P.O. Box 9513, 2300 RA, Leiden, The
Netherlands \\
\inst{2} Institute of Astronomy, Madingley Road, Cambridge CB3 0HA, United Kingdom\\
\inst{3} Joint Institute for VLBI in Europe, Postbus 2, 7990 AA, Dwingeloo
The Netherlands\\
\inst{4} Netherlands Foundation for Research in Astronomy, Postbus 2, 7990 AA, 
Dwingeloo, The Netherlands\\
\inst{5} Kapteyn Institute, Postbus 800, 9700 AV, Groningen, 
The Netherlands}

\maketitle
 
\markboth{Snellen et al.: Deceleration and the Morphologies of GPS sources}{}

\begin{abstract}
A relativistic radio component, which moves in a direction close to the sky
plane, will increase in flux density when it decelerates. This effect is the
basis for the qualitative model for GPS galaxies we present in this paper, 
which can explain their low-variability convex spectrum, their compact double 
or compact symmetric morphology, and the lack of GPS quasars at similar
redshifts. 
Components are expelled from the nucleus at relativistic speeds at a large
angle to the line of sight, and are decelerated (eg. by ram-pressure or
entrainment of the external gas) before contributing to a mini-lobe. The young
components are Doppler boosted in the direction of motion but appear fainter
for the observer. The non-relativistic mini-lobes dominate the structure and
are responsible for the low variability in flux density and the convex radio
spectrum as well as the compact double angular morphology. Had the same source
been orientated at a small angle to the line of sight, the young components
would be boosted in the observer's direction resulting in a flat and variable
radio spectrum at high frequencies. Hence the characteristic convex spectrum
of a GPS source would not be seen. These sources at small angles to the line
of sight are probably identified with quasars, and are not recognized as GPS
sources, but are embedded in the large population of flat spectrum variable
quasars and BL Lac objects. This leads to a deficiency in GPS quasars at
$z<1$, but the model does not explain the population of GPS quasars at high 
redshift. However, there is increasing evidence that the high 
redshift GPS quasars are not related to the GPS galaxies.
It is interesting for
unification purposes to investigate if deceleration is a general phenomenon in
compact radio sources. We have therefore investigated if evidence for
deceleration can be traced in flux density outbursts of highly variable radio
sources, and have found that indeed the evolution of the spectral peak in such
outbursts does suggest that deceleration may play an important role, at least
in some of these sources.
\end{abstract}

\section{Introduction}
Gigahertz Peaked Spectrum (GPS) sources are a class of extragalactic radio
source characterized by a convex radio spectrum peaking at a frequency of
about 1 GHz. They are compact luminous radio sources. The turnovers in their
radio spectra are believed to be due to synchrotron-self-absorption caused by
the high density of relativistic electrons in these compact sources. The
overall angular size of GPS sources is a function of their spectral
peak: the lower the peak frequency the larger the radio source. GPS sources may
therefore be related to the class of compact steep spectrum (CSS) sources,
which have lower peak frequencies ($\sim 100$ MHz) and larger angular and
linear sizes.

Optical identification programs carried out on GPS sources show the optical
counterparts to be a mixture of galaxies and quasars (O'Dea 1991, Stanghellini
et al. 1993, de Vries 1995). The quasars tend to have very
high redshifts, while the GPS-galaxies have redshifts which are generally well
below unity (O'Dea et al. 1991). Within the context of the simple orientation
unification scheme of radio galaxies and quasars, it is puzzling that the
redshift distribution of GPS galaxies is so different from that of GPS
quasars. There is an increasing amount of evidence available in the literature
that the GPS sources identified with galaxies and quasars are not related to
each other and are different classes of objects; in addition to their different
redshift distributions, it has been found that they have different
distributions in rest-frame peak frequency, linear size and radio morphology
(Stanghellini et al. 1996).

Early VLBI data on GPS sources showed that many of them, especially those
identified with galaxies, had a compact double (CD) morphology with components
of similar flux densities and spectral indices (Phillips and Mutel 1982). This
led to the speculation that these are young radio sources, in which the two
components could be interpreted as being mini-lobes. More recent higher
quality images show that in many CDs weak central components with flatter
spectral indices are found which are most naturally interpreted as being the
central core. These sources were renamed compact symmetric objects (CSO), and
 it has been proposed
that CSOs evolve into FRI and/or FRII sources
(Readhead et al. 1994, Fanti et al. 1995). Stanghellini et al. (1996) has
recently shown that about 90\% of the GPS sources identified with galaxies are
CSOs.

In this paper we present a model for GPS galaxies which can explain their
spectral shape, their compact double or compact symmetric morphology, their
apparent low flux density variability and the lack of GPS quasars at similar
redshifts. In section \ref{sec21}, the observational basis for the model is
presented. In section \ref{sec22}, the model is explained and its implications for the
appearance of GPS sources presented. 
In section \ref{sec3}, the possiblity that deceleration is a more general phenomenon
is investigated, by studying the spectral evolution of flux density outbursts
in highly variable radio sources.

\begin{figure}[!t]
\centerline{
\psfig{figure=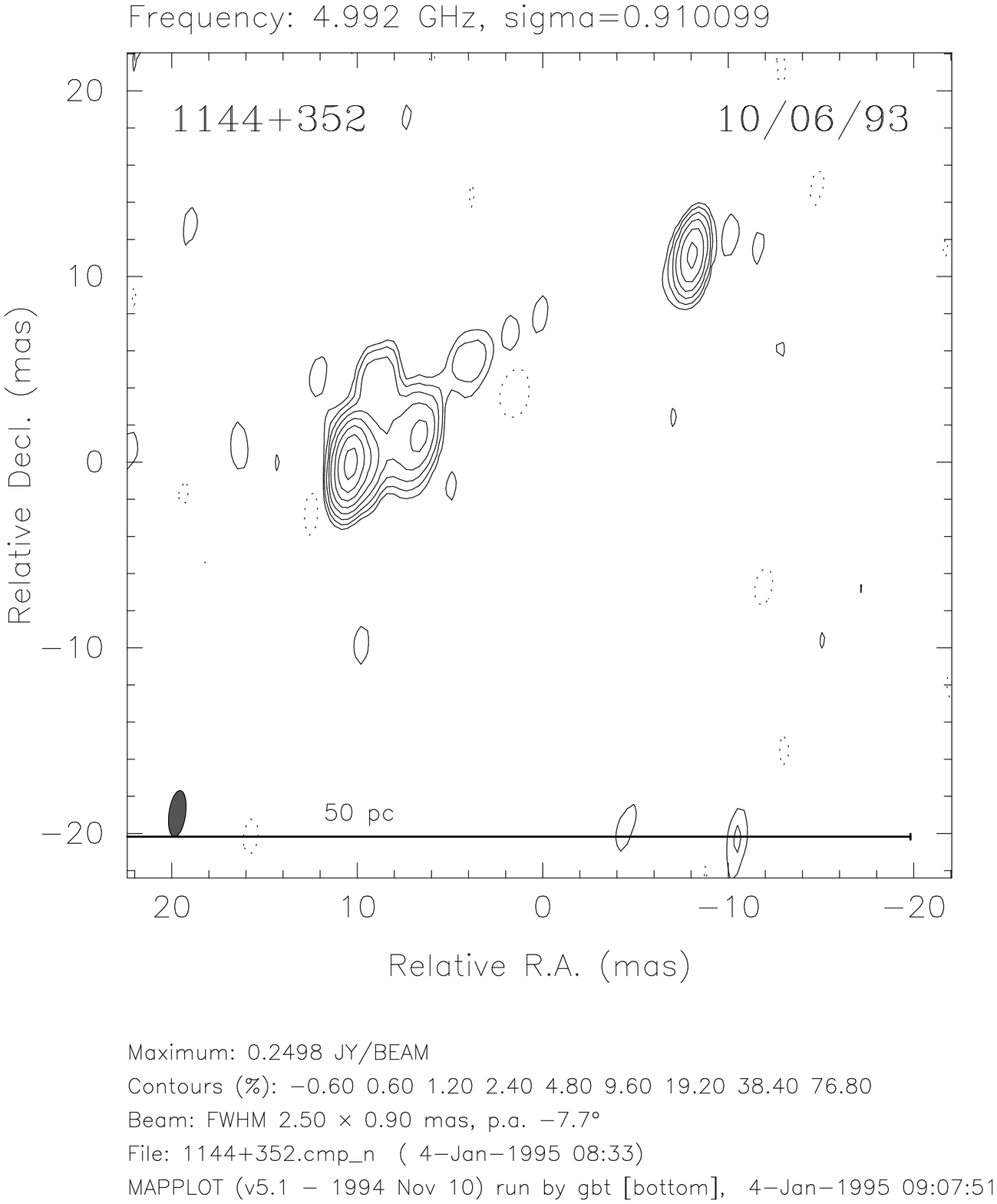,bbllx=25pt,bblly=120pt,bburx=587pt,bbury=720pt,clip=,width=8.5cm}
}
\caption{\label{1144} 1144+352 at 5 GHz from the Caltech-Jodrell II survey 
(Henstock et al. 1995). The
brighter component to the east has a flux density of 467 mJy and is mainly
responsible for the observed flux variability over the last 20 years. The two
brightest components separate from each other with an apparent velocity of
$1.2c/h^{-1}$ (Giovannini et al. 1995). }
\end{figure}

\begin{figure}[!t]
\centerline{
\psfig{figure=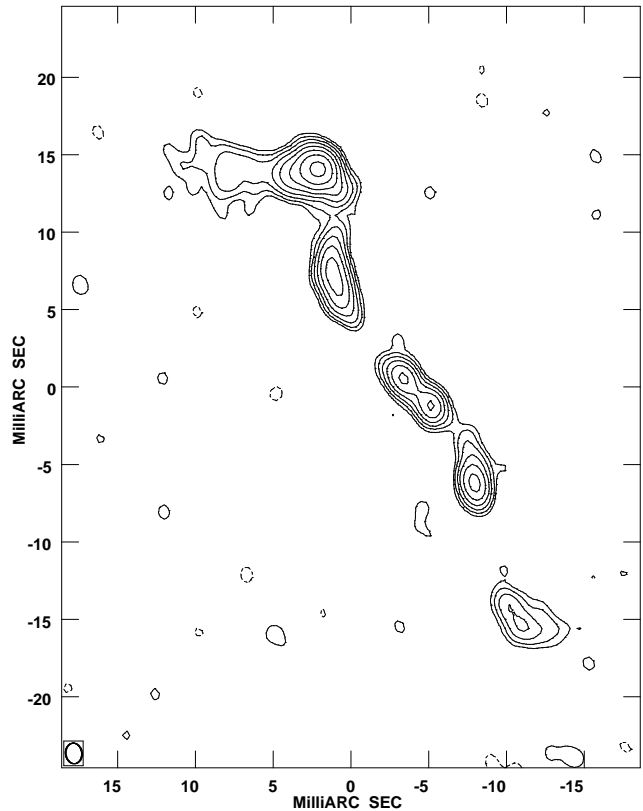,bbllx=51pt,bblly=90pt,bburx=560pt,bbury=734pt,clip=,width=8.5cm}
}
\caption{\label{1946} 1946+708 observed at 5 GHz by Snellen et al.
(in prep.).
Taylor et al. (1997) found bi-directional motion for the inner components, 
while for the outer components no motion is detected. This indicates that the 
components are decelerated while moving outwards.}
\end{figure}

\section{ The Spectral and Angular Morphology of GPS Galaxies}

\subsection{\label{sec21} The Starting Points}

The following three observations of properties must be explained by any 
model of GPS sources:
\begin{itemize}
\item[1]  {\bf The symmetric structures of GPS sources.}
The compact symmetric structure seen in most GPS galaxies 
(Stanghellini et al, 1996) indicates that 
relativistic boosting is relatively unimportant, otherwise only radio emission
on the side directed towards us would be visible.
 This implies that these components are not ejected from the nucleus at
small angles to the line of sight and at highly relativistic speeds.
However it is not excluded that the components move at highly relativistic 
speeds at large angle to the line of sight (see point 2).
\item[2] {\bf The relativistic velocities of components in nearby GPS galaxies.}
In two nearby GPS sources, 1144+352 (Giovannini et al. 1995)  and 1946+708
(Taylor et al. 1997), it has been shown that some components have relativistic
motions. The source 1144+352, which is shown in figure \ref{1144}, consists of
two components (ignoring faint large scale structure at long wavelengths)
which separate from each other with an apparent velocity of $1.2c/h^{-1}$. 
The structure of 1946+708 consists of a series of components (see figure 
\ref{1946})
found by Taylor et al. to move away bi-directionally from a point between
the two central components, probably the location of
the nucleus. It is noteworthy that the components close to the centre move
outwards faster than the components further away and that no motion is found
for the two components  at the largest distances from the centre. The highest
apparent velocity measured  in this source is $0.9c/h^{-1}$ (Taylor et al
1997).
\item[3] {\bf The flux density variations of the nearby GPS source 1144+352.}
Although the GPS source 1144+352 is variable on timescales of several years 
(Snellen et al. 1995), its
peaked spectral shape does not change. The peak flux
density of the source increased monotonically from 300 to 600 mJy over some 20 years,
and is at the moment decreasing slowly again. In view of the simple structure
of 1144+352, we can attribute this flux density variation completely to the
eastern-most (brightest) component in this source (Giovannini et al. 1995),
because it contributes 85\% of the flux density at 5 GHz. 
\end{itemize}

The variability of 1144+352, the relativistic velocities of components in
nearby GPS sources, and the large angles to the line of sight implied by
symmetric morphologies lead us to the model discussed below.

\subsection{\label{sec22} A Decelerated Component Model}

\begin{figure*}[!t]
\psfig{figure=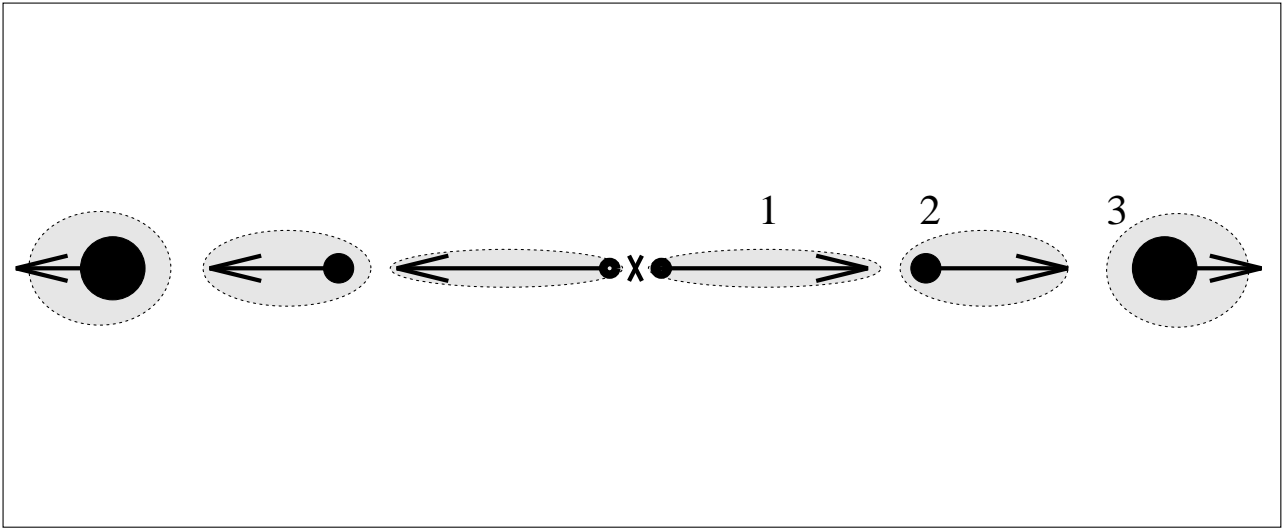,width=17cm,angle=-90}
\caption{\label{model1} The decelerated component model. The cross indicates the 
central engine. Components are expelled 
from the nucleus at relativistic speeds (1), then decelerate (2) and then
contribute to a mini-lobe (3). The arrows indicate the velocity, and the 
grey ellipses the anisotropy of the observed radio emission due to 
Doppler boosting.}
\end{figure*}

We shall model a GPS radio galaxy as a series of radio components 
expelled at different epochs, hence observed at different stages of their 
evolution. 
Each of those components is assumed to be expelled from the nucleus at
relativistic speeds along a direction which makes a large angle to the line
of sight. These components then decelerate and expand as they interact with the
surrounding medium.
In section \ref{nature} we will go briefly into the possible nature of the expelled
components and their deceleration mechanism, however this is not necessary 
for our qualitative approach and not important at the moment.

Consider the evolution of one such  component. Initially its velocity is highly 
relativistic (figure \ref{model1}, component 1) and the intensity of its radio emission
is strongly Doppler boosted in the direction of motion. The component is 
assumed to be sufficiently compact to be optically thick at low frequencies
as a result of synchrotron self absorption (SSA).
Assuming that the component emits isotropically with a flux
density $S_0(\nu )$ at frequency $\nu $, it will be observed to have a flux
density of
\begin{equation}
S(\nu \delta )=S_{0}(\nu) \delta ^3
\end{equation}
where $\delta$ is the Doppler factor, $\delta=(\gamma (1-\beta \cos{\theta
}))^{-1}$, $\beta$ is the velocity in units of the velocity of light, $\gamma
= 1/\sqrt{1-\beta ^2}$, and $\theta $ the angle of the direction of the motion
to the line of sight. 
For small values of $\theta $, and  highly relativistic
velocities this results in $\delta >> 1$. 
However, for a large angle to the
line of sight, $\cos{\theta }$ will be small, hence $\delta \approx \gamma
^{-1}$, which results in $\delta < 1$. As it evolves, the component will be
decelerated to a modest relativistic speed (figure \ref{model1}, component 2), as is seen
in 1946+708, eg. by drag or entrainment of the external gas. 
The component 
also expands,  resulting in a intrinsically lower SSA peak frequency and peak
flux density.

\begin{figure}[!t]
\centerline{
\psfig{figure=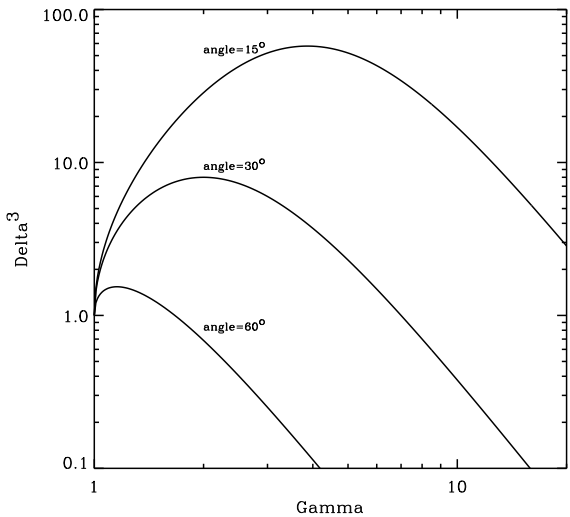,width=8cm}}
\caption{\label{boosting} The Doppler factor $\delta$ to the third power, which is 
proportional to the peak flux density of a source, as function of
$\gamma$ for an angle of $15^{\circ}$, $30^{\circ}$ and $60^{\circ}$ 
to the line of sight.}
\end{figure}

Figure \ref{boosting} shows the Doppler factor as function of 
$\gamma$ for different angles to the line of sight.
For small $\theta $s and decreasing $\beta$, the Doppler factor decreases 
because
$\frac{1}{1-\beta \cos{\theta }}$ decreases faster than $1/\gamma $ increases.
However at large angles to the line of sight, $\delta$
increases with time. 
For $\theta = 90^{\circ}$, the peak flux density, which is
proportional to $\delta ^3$, increases by  a factor of 10 between $\beta =
0.98$ ($\gamma = 5$) and $\beta = 0.90$ ($\gamma = 2.3$). 
Eventually
the component is decelerated to barely relativistic speeds or slower and
will contribute to a 'mini-lobe' (figure \ref{model1}, component 3).

\begin{figure*}[!t]
\psfig{figure=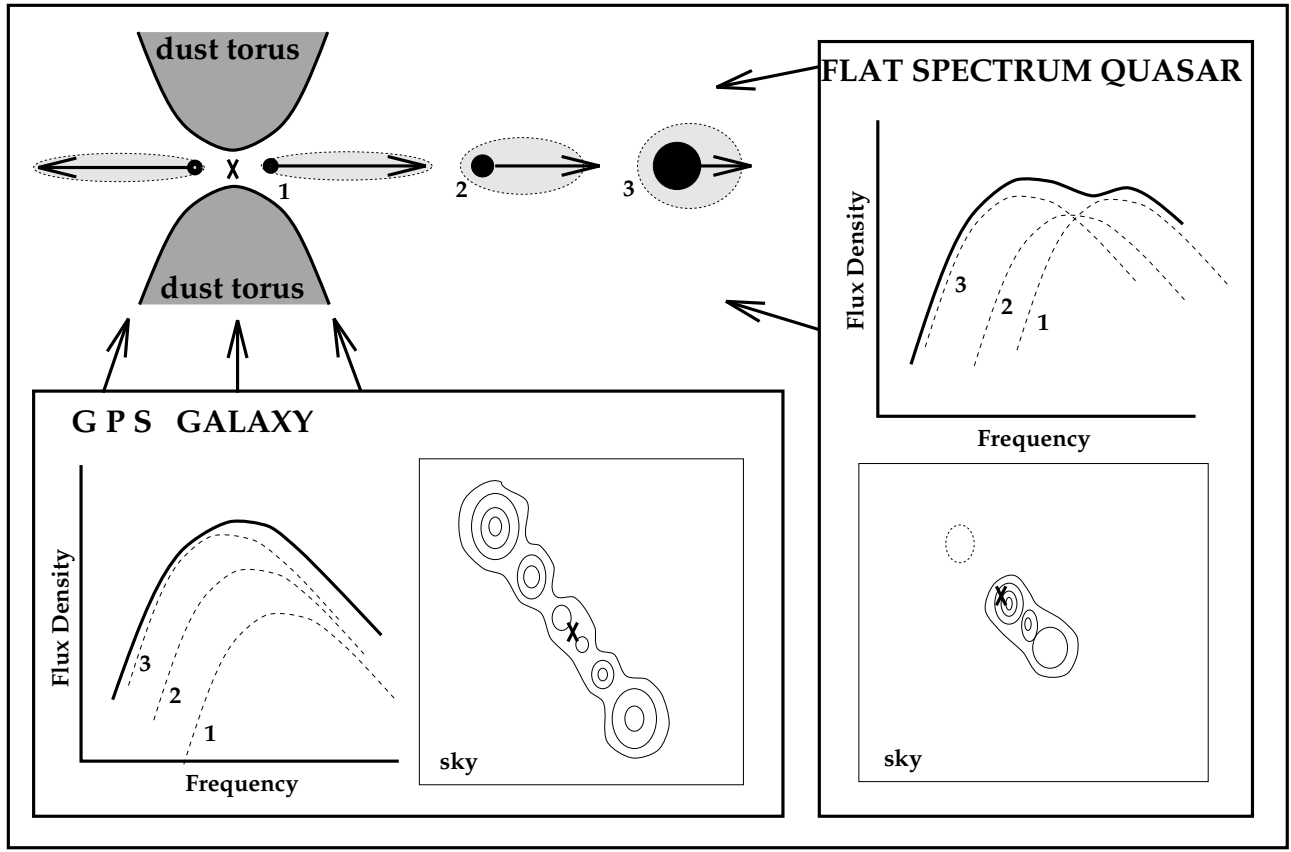,width=17cm,angle=-90}
\caption{\label{model2} The decelerated component model combined with the orientation
unification scheme. At a large angle to the line of sight, the Doppler boosting
of component 1 does not make a significant contribution 
to the overall radio spectrum, resulting in a GPS spectral morphology.
In this case the object is seen as a galaxy.
At a small angle to the line of sight, the young components moving towards
us are strongly Doppler boosted, resulting in a one sided jet morphology
and a variable flat spectrum. In this case the object is seen as a quasar.
Hence the lack of GPS quasars at similar redshifts to the GPS galaxies 
is due to their not being recognized as GPS sources, since they are embedded
in the large population of variable quasars and BL Lac objects, with radio 
emission only at sub-arcsecond scale.}
\end{figure*}

We now consider the effect of the simultaneous presence of
the fast and slow components on the overall
angular and spectral morphology that would be observed in a
population of radio sources (figure \ref{model2}). 

i) For objects whose radio axes are oriented at
a large angle to the line of sight, the slow, barely- or
non-relativistic components, which form mini-lobes on both sides of the
nucleus, are dominant in both the radio spectrum and angular morphology. The
mini-lobes are still compact enough to be self absorbed, resulting in a
'gigahertz peaked spectrum', and the radio source will have a compact
symmetric morphology. The fact that $\delta < 1$ for the highly relativistic
young components, makes them appear faint. Therefore the flux density 
variability will be
low due to the small contribution to the total flux density of the young and
fast components; that is the creation of a new component will only have a small
impact on the overall radio spectrum.

ii) For objects whose radio axes are oriented at
 a small angle to the line of sight, the influence of Doppler boosting on the
appearance of the (slow) mini-lobes is small. They still would be
observed to have a `gigahertz peaked' radio spectrum, were it not that the
young components, which move towards the observer, are strongly boosted in
flux density and blue-shifted in frequency and dominate the overall spectrum
producing a peak at high frequency. The overall radio spectrum appears to be
flat or inverted with a peak at high frequency (see figure \ref{model2}), and 
the object
is not classified as a GPS source. Furthermore, the creation of new components
produces a large effect on the radio spectrum, and results in the source being
observed as (strongly) variable at high frequencies.

According to the orientation unification scheme 
(eg. Barthel, 1989) a quasar whose radio axis is oriented at a large 
angle to the line of sight, is observed to be a radio galaxy.
However in our model, a GPS galaxy observed at a small angle to the line of
sight, would not result in a quasar with a GPS radio spectrum, but in a 
quasar or blazar with a flat or inverted variable spectrum. 
Therefore no population of GPS quasars at similar redshifts to 
the GPS galaxies will be observed. These quasars would hide as a subsample in 
the large population of flat spectrum quasars and BL Lac objects.

\section{\label{sec3} The Evolution of a Single Component}

The physical properties of radio jets are widely discussed in the
literature, but this has not yet converged to a single view.
Fortunately, no detailed understanding of the underlying jet-physics is 
necessary to come to a qualitative description of the flux density evolution
of a decelerating radio component. However, we will discuss briefly 
the possible nature and deceleration mechanisms of the radio components
in GPS sources.

\subsection{\label{nature} The Physical Nature of the Radio Components and 
the Deceleration Mechanism.}

In general it is assumed that radio jets can be treated as continuous
fluids, but the 
jet-physics of GPS sources may be different from radio sources in general 
due to their young age; eg. a steady continuous jet may not have been formed 
yet.
The most simple discrete radio source model is the adiabatic, spherically 
expanding
source model (eg van der Laan 1966, Pauliny-Toth and Kellermann 1966), which
was the first quantitative  attempt to explain radio outbursts.  
If the radio components could indeed be treated as individual `plasmons' the
deceleration mechanism can be easily explained due to ram-pressure of the
surrounding medium. Such a scenario has been modeled by Christiansen et al.
(1978). However, this scheme results in a large energy release by
dissipation which should be observed in some way.

In radio source models which consider jets to be continuous, the radio
components are in general treated as shocks. In this case deceleration
of the jets can be caused by entrainment of the external gas. When
deceleration of jets was first discussed for FRI radio sources
(eg. Begelman 1982, Scheuer 1983), it was pointed out that if an
initially relativistic jet decelerates to sub-relativistic speeds much
of its kinetic energy is thermalized, but that the jet can be
recollimated by an external pressure gradient. More detailed analyses
have been carried out by Bicknell (1994, 1995), Komissarov (1994).
Bowman et al. (1996) claim that the energy released by dissipation is 
re-converted continuously into kinetic energy since the jet is propagating
down a pressure gradient, and that deceleration without catastrophic 
kinetic energy loss is relatively easy to achieve for relativistic jets
compared to non-relativistic ones.
  Recently, de Young (1997) concluded that for a
constant hot-spot advance speed for CSOs as derived from VLBI data by Readhead
et al. (1996) it is necessary to have entrainment along most of the
jet, therefore allowing deceleration of the jet.

\subsection{The Qualitative Flux Density Evolution of a Decelerating
Component}

If the deceleration model described above is valid, then deceleration should 
have a large
influence on the spectral evolution of flux density outbursts observed in
quasars and BL Lac objects. It is therefore interesting to investigate these
flux density outbursts, and see if there is any signature of
deceleration. First of all we have to discuss
how flux density outbursts are explained in current models.

\begin{figure}[!t]
\hspace{-0.2cm} 
\centerline{
\psfig{figure=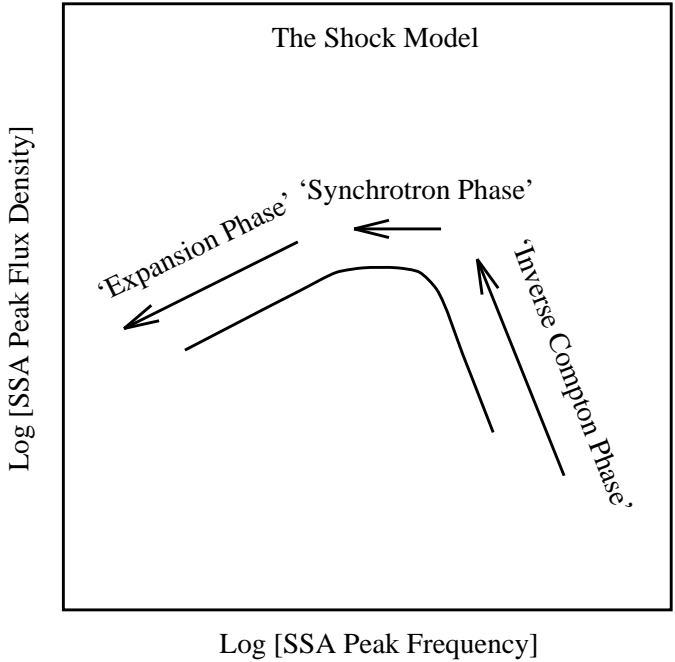,width=9cm,angle=-90}
}
\caption{\label{shock} The evolution of the SSA spectral peak of a radio 
outburst according to the shock model (Marscher and Gear 1985). Note that in 
the first phase, where inverse Compton losses are dominant, the
peak frequency is expected to decrease with time.}
\end{figure}

The adiabatic expanding source model agreed well with the observed 
spectral evolution at low frequencies, but it was less successful in 
explaining the high-frequency behaviour due to the earliest stages of the
outbursts. Closer agreement was obtained using shock models,
in which the radio outbursts are explained as due
to propagation of shocks in relativistic jets (Marscher and Gear 1985, Hughes
et al. 1989). The general applicability of the shock models is now widely 
accepted, although
there are some features that these models cannot yet explain properly (eg. the
growth stage of the shock). Three 
phases can be distinguished in the shock model (see figure \ref{shock}). In the
first phase, Compton losses are dominant, resulting in a dependence of the SSA
peak flux density $S_{peak}$ to the peak frequency $\nu _{peak}$ of the
shocked component given by $S_{peak} \propto \nu _{peak} ^{-5/2}$. In the second
phase, synchrotron losses are assumed to be dominant, resulting in $S_{peak}
\propto \nu _{peak} ^{-0.05}$. In the last phase, expansion losses are dominant,
which is comparable to the adiabatic expanding radio source model, resulting
in $S_{peak} \propto \nu _{peak} ^{+0.5}$. Note that the exact relations between
$S_{peak}$ and $\nu _{peak}$ in this model are both dependent on the
relativistic electron  energy distribution and the change in magnetic field
strength along the jet (see Marscher and Gear 1985). 
The predictions of
this shock model agree well with spectral observations (eg. Lainela 1994).

\begin{figure}[!t]
\hspace{-0.2cm} 
\centerline{
\psfig{figure=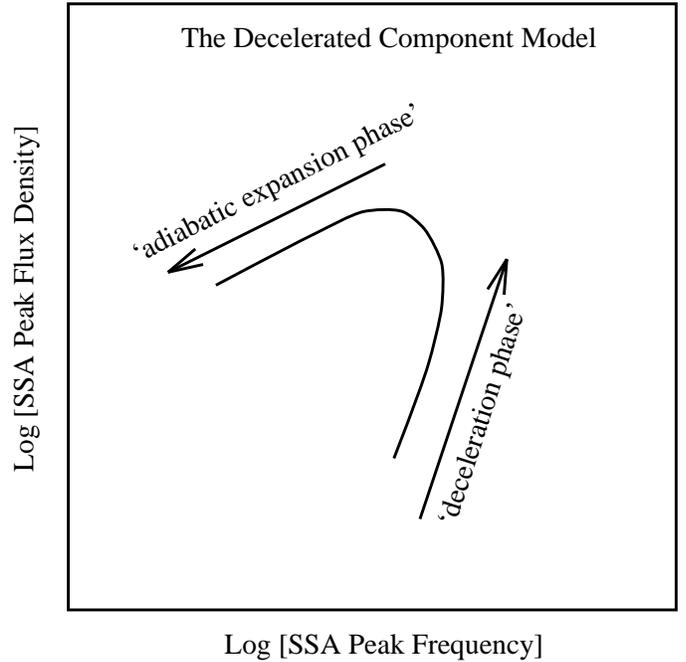,width=9cm,angle=-90}
}
\caption{\label{decel} The evolution of the SSA spectral peak of a radio outburst 
according to the decelerated component model. Note that in 
the first phase, where deceleration is dominant, the
peak frequency is expected to increase with time.}
\end{figure}

What influence can deceleration have on the spectral evolution of a radio
outburst? Assuming  that deceleration is the dominant factor,
the change in the Doppler factor $\delta$ results in a
change of the SSA spectral peak. 
From equation 1 we find that $\Delta S_{peak} \propto (\Delta\delta)^3$ and
$\Delta \nu_{peak} \propto \delta$, and therefore that
$\Delta S_{peak} \propto
(\Delta \nu _{peak})^{3}$. If the velocity of the component is highly 
relativistic, and
$\gamma >>1$, deceleration will cause an increase in $\delta$, resulting in
an increase in peak flux density and peak frequency, even for a small angle
to the line of sight (see fig. \ref{boosting}). 
For example, at an angle of $20^{\circ}$ to the line of
sight, $\delta$ is maximized at $\gamma$=2.9. Hence for
$\gamma > 2.9$ deceleration will cause an increase in $\delta$. Between
$\gamma = 10$ and $\gamma = 2.9$, the observed peak flux density will increase
by a factor of 40, and the peak frequency increases by a factor of 3.4. 
During the deceleration of a component, it will also expand. At a certain
point, the change in $\delta$ will be small compared to the changes
in the spectrum due to the expansion. 
This period will be similar to the last phase in the general shock model.
The effect of combined deceleration and adiabatic expansion is shown in 
figure 7,
in which it is clear that the predicted spectral evolution is comparable
to that expected for the generalized shock
model, except in the initial phase of the outburst. The deceleration phase
in fact replaces the inverse Compton and synchrotron phases.
How can we distinguish between the two models?
A detection of an increase in peak frequency combined with an increase in
peak flux density of a flux density outburst would underline the
importance of deceleration.
However, this is difficult to observe, because it is expected to occur 
in the very early phase of the outburst.
We have searched the literature for examples, as we now describe.

\subsection{Observations of the Spectral Evolution of Flux Density Outbursts.}

To be amenable for investigation of its spectral evolution, a radio outburst 
should have properties which obey the following criteria:
\begin{itemize}
\item[1)] The outburst should be a single event uncontaminated by
previous outbursts. Blending of spectral features due to different outbursts
make detailed investigation impossible.
\item[2)] The radio spectrum during the initial period of the outburst, 
before maximum flux density is reached, must be well sampled.
This implies that the timescale of the outburst should be sufficiently long for
this to be accomplished.
\item[3)] The radio source must be monitored at three or more frequencies to
obtain sufficient information about the evolution of the spectral peak.
Furthermore the highest frequency at which the source is monitored should be
in the optically thin part of the spectrum, to allow the peak to be properly
fitted. This implies that the maximum flux density of the outburst at the
highest observed frequency should be lower than the maximum flux density at
lower frequencies.
\end{itemize}  

\begin{figure*}[!t]
\hbox{
\hspace{-1cm} \psfig{figure=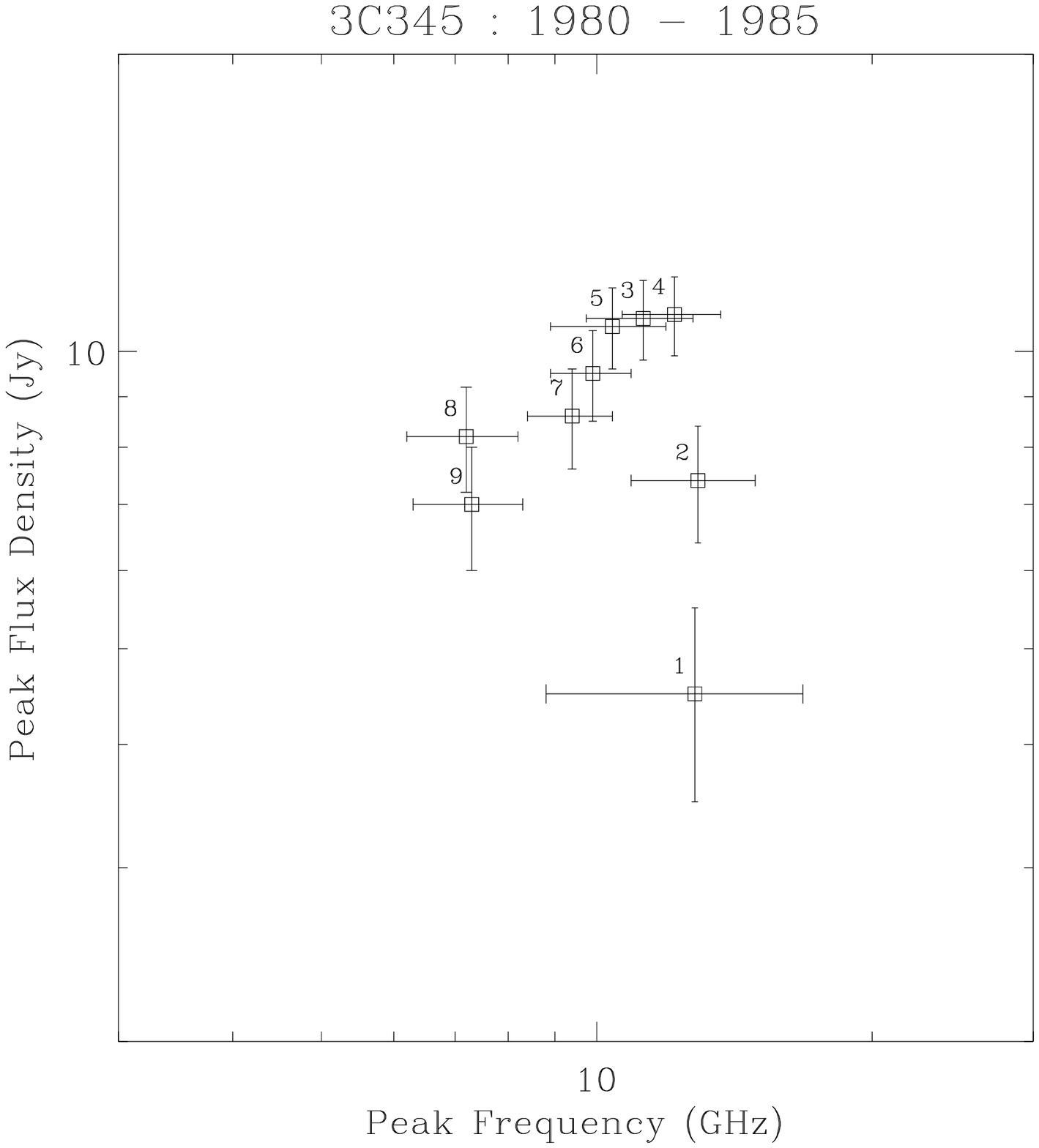,width=6.2cm}
\hspace{-1cm} \psfig{figure=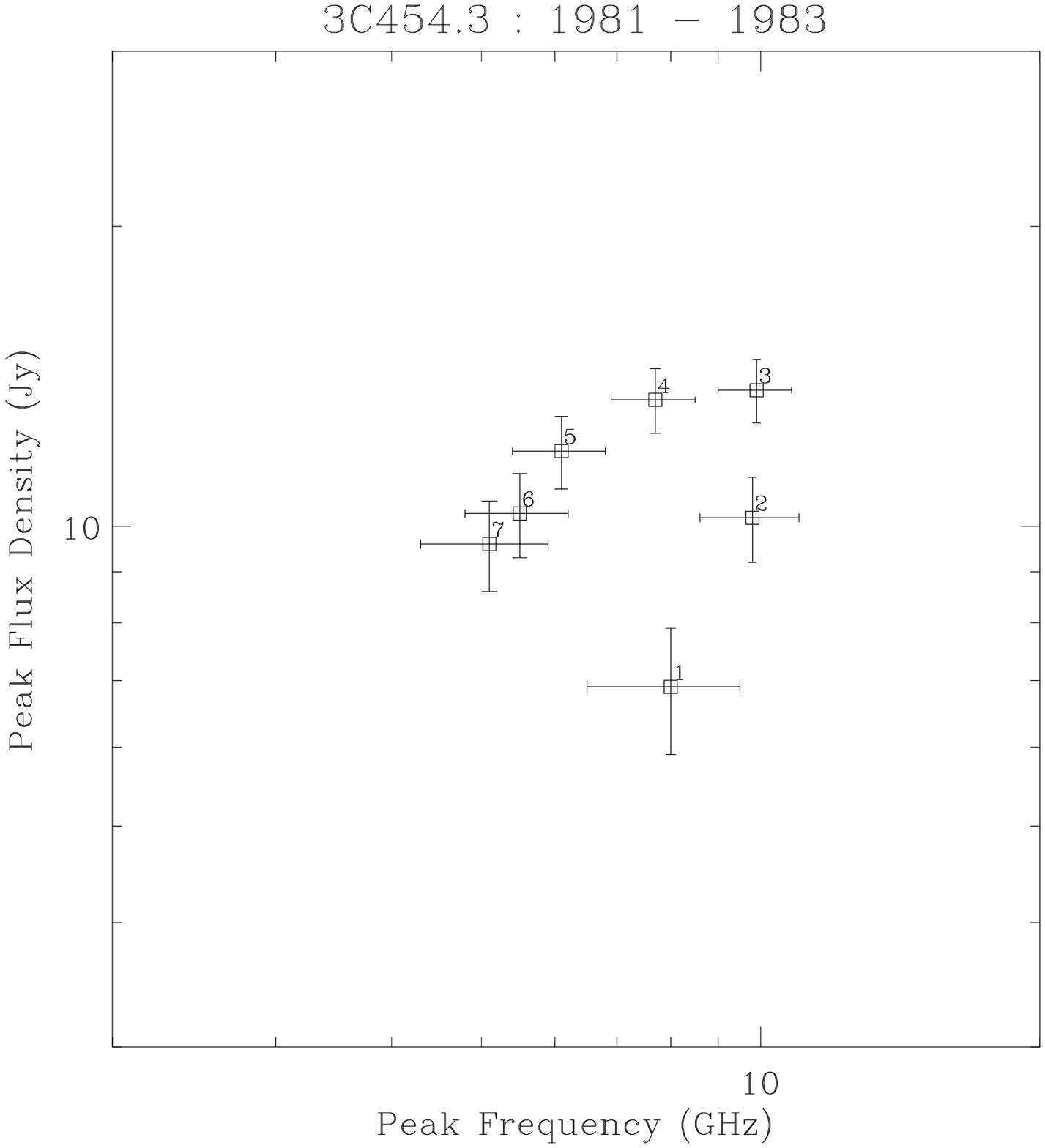,width=6.2cm}
\hspace{-1cm} \psfig{figure=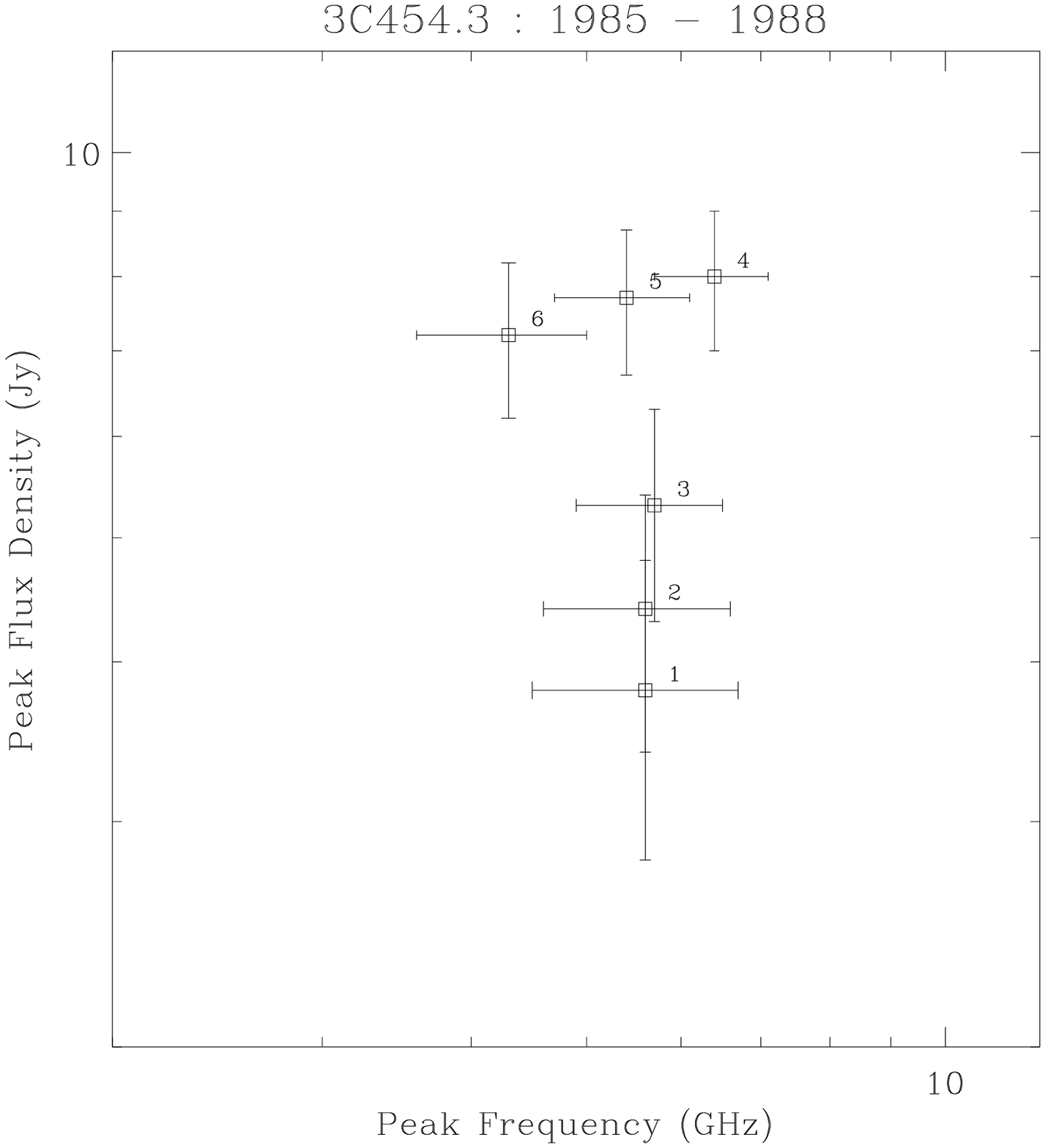,width=6.2cm}}
\caption{\label{outburst} The evolution of the spectral peak during outbursts in 
3C345 (left) and 3C454.3 (middle and right). The numbers indicate the 
dates of the observations. The dates for 3C345 are 1980.5 and every half
a year after. The dates for 3C454.3 are 1981.1 and every 0.3 years later for
the first outburst, and 1985.5 and every half year later for the second 
outburst. Note that, especially the 1982 outburst of 3C454.3, undergoes an
increase in peak frequency during the first stage of the flare which is not
expected for the generalized shock model. However this is expected if
deceleration is important.}
\end{figure*}

For many years large dedicated monitoring programs 
to study the time variability of radio sources have been carried out by 
the University of Michigan at 4.8, 8.0 and 14.5 GHz (Aller et al. 1985)
and the Mets\"ahovi Observatory at 22 and 37 GHz (Salonen 1983, 1987, 
Ter\"asranta 1992).
We used data obtained in the course of these two monitoring programs to search 
for radio outbursts with properties obeying the criteria mentioned above.
The best cases we found are the outbursts in 3C345 and 
3C454.3 
peaking in 1981 and 1982 respectively. Furthermore, Margo Aller provided
us with unpublished data from an outburst in 3C454.3 in 1986.
All three outbursts have well sampled observations at the 5 frequencies
mentioned  above.
The contribution of emission not related to the outburst,
often refered to as quiescent emission, was estimated from the flux 
densities in the period before the outburst, and then subtracted from
the total observed flux densities. 
For 3C345 the quiescent level is about 7.0 Jy at all observed frequencies, 
and for 3C454.3 the quiescent emission was estimated
to be $S_q = 10.0\nu ^{-0.2}$.
The resulting spectra were fitted with a least squares fitting routine, in
which  
the optically thin spectral index was fixed at $\alpha =-0.75$. Note that if the
optically thin spectral index were to change during the outburst, it would become
steeper in time (due to aging of the high energy electrons) and results in an 
estimation of the peak frequency which is too low. Therefore
keeping $\alpha$ constant cannot be the reason for finding an
increase in peak frequency with time. The resulting evolution of the 
spectral peak
in time for the three different outbursts are shown in figure 7. In the 
outburst in 3C345 and the second outburst in 3C454.3, the peak frequency does 
not significantly change in the initial stage of the outburst. However the 
first
outburst of 3C454.3 seems to have an increase in peak frequency in the initial
stage of the outburst, which is expected if deceleration plays an important
role. In all three cases the peak frequency to peak flux density relation is
steeper than expected for the general shock model 
($S_{peak} \sim \nu _{peak}^{\kappa }$, with $\kappa = -5/2$), 
we find $\kappa = -6.7$ for the 3C345 outburst and $\kappa = +1.9$ and $+3.6$
for the 3C454.3 outbursts.
Hence these observations support the idea that deceleration plays a
role in the spectral evolution of radio outbursts and, in some cases,
may have a dominant role, e.g. 3C454.3.

Another method of investigating the spectral evolution of a radio source 
during an
outburst would be to monitor its spectral index between two frequencies far 
enough
apart that the spectral peak is enclosed between them. 
This has been done by Stevens
et al. (1995) for the blazar PKS 0420-014. They have monitored the spectral
index between 90 and 22 GHz and find  clear relations between the 90 GHz
flux density and the spectral index for the rising and
falling phases of its radio flux density evolution. 
These relations can be converted to a relation between peak
flux density and peak frequency for both stages, assuming the quiescent
level (given by Stevens et al.) and a homogeneous self absorbed source with
an optically thin spectral index, for which
we take the value of Stevens et al. of -0.44 and assume it to be constant.
During the rise in flux density the evolution of the spectral peak obeys the
relation $S_{peak} \propto \nu _{peak}^k$, where $k=+2.0^{+1.7}_{-0.7}$. During
the decay in flux density it follows the same relation with $k=+1.1\pm0.1$.
Hence these observations also support the idea of deceleration.
  
\section{Discussion}

\begin{figure}[!t]
\hspace{-1cm}
\centerline{
\psfig{figure=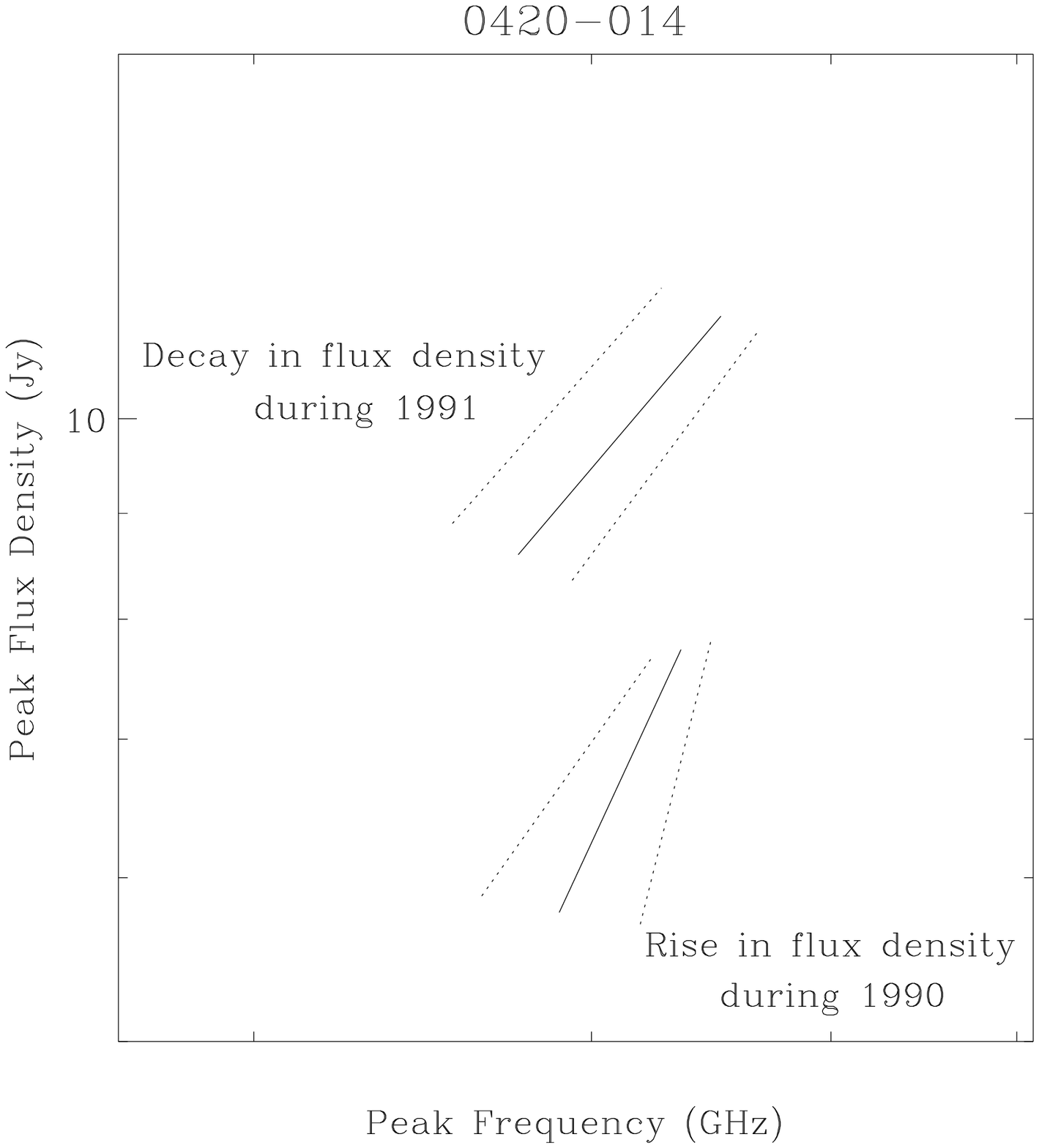,width=9cm}
}
\caption{\label{0420} The evolution of the peak frequency and peak flux density of a
flux density outburst of 0420-014, derived from the spectral index monitoring
by Stevens et al. (1995) at 90 and 22 GHz, assuming a homogeneous self-absorbed
source. During the rise in flux density the evolution of the spectral peak
the peak frequency increases, hence these observations also support the idea 
of deceleration.}
\end{figure}

The decelerated component model which we discuss in this paper can explain
several properties of Gigahertz Peaked Spectrum radio galaxies. While  
consistent with the typical two-sided morphology and low variability,
it can also explain the lack of GPS quasars at similar redshifts as the 
GPS galaxies, and account for the
spectral evolution of flux density outbursts in highly variable radio sources. 
In this scheme, 1946+708, our prototype compact symmetric GPS source, is
oriented close to the plane of the sky and
would not have had a gigahertz peaked spectrum, if it were oriented at a small
angle to the line of sight. It has a peak frequency of about 2 GHz and a 
peak flux density of about 1 Jy, and is believed to lie at a large angle to 
the line of sight, $\theta = 75^{\circ}$ (Taylor et al. 1995). 
The measured flux density of the innermost component is about 50 mJy 
at 5 GHz. If the angle to the line of 
sight had been $30^{\circ}$, and assuming that its velocity is 0.9c, the 
doppler factor had been a factor 3.5 higher and
this young component would contribute 2 Jy at 17.5 GHz. 
Hence the overall spectrum of this radio source would appear 
flat above 1 GHz and it would not be classified as a GPS source. 
Furthermore the rapid change in Doppler boosting and 
expansion of the young component would make the radio source appear 
significantly 
more variable. For objects whose radio axes are oriented
at small angle to the line of sight, the young components
will be dominant at high frequencies, but at about 1 GHz, they would be
observed in 
the optically thick part of their spectra where Doppler boosting is only 
mildly important ($S \propto \delta^{0.5}$). Hence their 1 GHz flux densities
would be similar to those observed if they were oriented
at a large angle to the line of 
sight. 
A GPS galaxy observed at $\theta <20^\circ$ will appear as a flat spectrum
variable quasar or a Bl Lac object. In a flux density limited sample at low 
frequency, only 1 flat spectrum quasar  with no extended emission is
expected per 20 GPS galaxies assuming a parent population with a random
distribution of $\theta$.

This corresponds to only 1 to 2 objects with a flux density greater than 1 Jy 
at 1 GHz.
An example of such an object could be PKS 1413+135 (Perlman et al. 1994),  
which has no measured extended emission, a ``flat'' variable spectrum at
high frequency, and at VLBI scales a component opposite to the jet 
which can be interpreted as a non-Doppler-boosted mini-lobe on the
receding side of the source. Another candidate GPS source seen at 
a small angle to the line of sight is 1504+377 (CJI Survey,
Polatidis et al. 1995). 

Although the angle to the line of sight has to be large to produce a Gigahertz
Peaked Spectrum, $\theta$ will not often be exactly $90^{\circ}$. Therefore one
side will have a component directed towards us and the other side will
have a component directed away from us.
This will result in
different Doppler boosting for the approaching and receding jet, especially
for the inner part of the jet which is moving the fastest. 
Such a situation could account for the morphologies observed in
 sources like 0108+388, 0404+768, 2352+495 and 0710+439 (Taylor et
al. 1996): While the outer components are comparable in flux density, the inner
components are only visible in the approaching jet. An exception to this
is our prime example 1946+708. The inner two components are comparable in flux
density for both the approaching and receding jet. This could be due to the
fact that we observe these components in the optical thick, inverted parts of 
their spectra which are not very sensitive to Doppler boosting. 
However the outer two
components, which show no motion, differ by a factor of $\sim 5$ in flux
density. This could be caused by inhomogeneity of the medium for example, 
which is proposed by Taylor et al. 1995. 
However, there is a possibility that due to the fact that the
approaching component is observed at an intrinsically later epoch than the
receding component, that this difference in flux density is a result of
observing the components at different evolutionary stages. VLBI observations
at different frequencies are needed to further investigate this.

So far, we have ignored the Gigahertz Peaked Spectrum radio sources identified
with quasars, which are found at high redshift. 
The model does not explain their existence. However, the GPS 
radio galaxies and quasars have a different distribution in redshift, 
rest frame peak frequency, linear size and VLBI morphology, and are
probably different classes of objects (Stanghellini et al. 1996).
Still it is confusing that these quasars, which are believed to be observed at
a small angle to the line of sight, do not undergo significant Doppler boosting.

\begin{figure*}[!t]
\psfig{figure=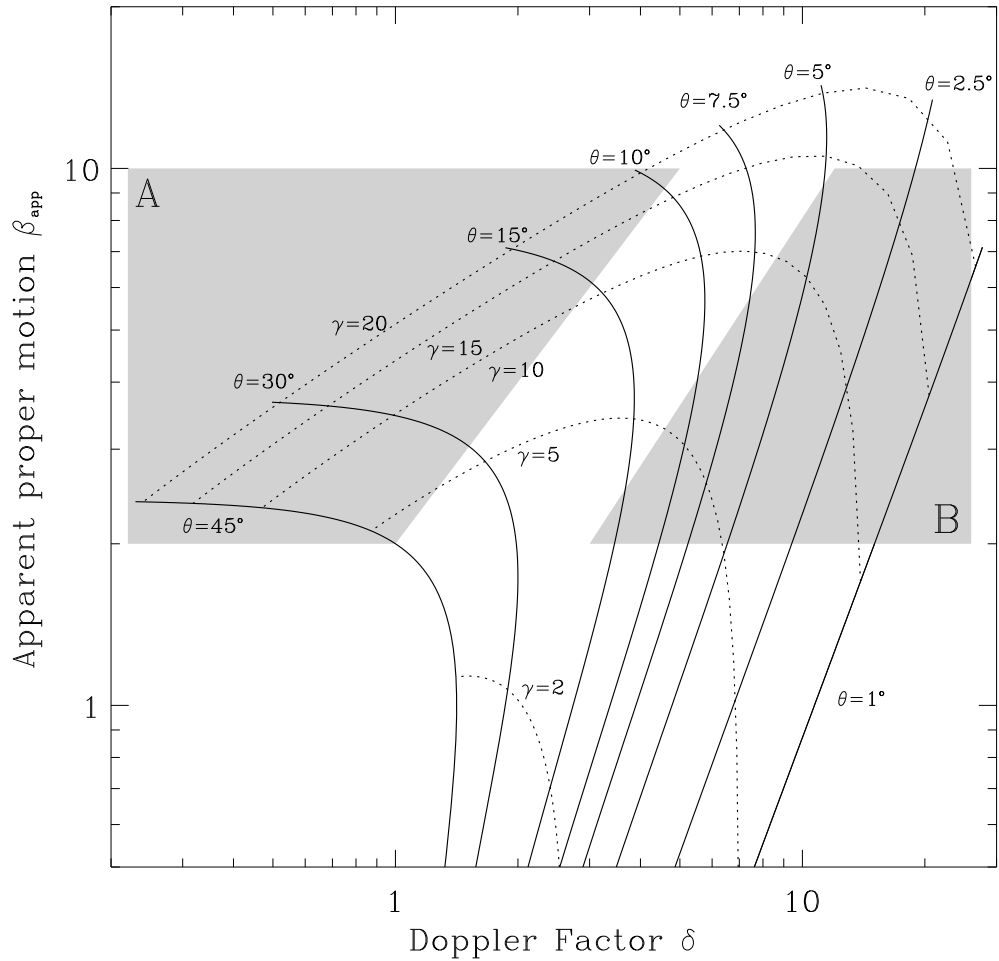,width=15cm}
\caption{\label{deltabeta} The influence of $\gamma$ and the angle to the line of sight $\theta$
on the Doppler factor $\delta$ and the apparent proper motion $\beta_{app}$.
The solid lines represent a range of $\gamma$s for a certain fixed $\theta$,
while the dotted lines represent a range of $\theta$s for a certain $\gamma$.
The areas $A$ and $B$ indicate the parameter space for which
an increase in $\delta$ can be due to deceleration
and acceleration respectively.} 
\end{figure*}

We have shown that deceleration could play an important role in 
3C454.3 and 0420-014 and have hypothesized that it plays an important role
in GPS galaxies and in other flat spectrum variable quasars and BL-Lacs.
We also have noted that there is in fact very little data available in which
the spectral evolution of individual outbursts can be unambiguously followed
to look for evidence of deceleration.
However it remains true that these outbursts can be exceptions. 
Furthermore, the
quiesent radio flux density was assumed to be constant, and the change in 
radio flux density is all contributed by one component. 
This does not have to be the case,
and the influence of other components can change the observed relation
between peak flux density and peak frequency. 

Ideally, one should analyse an outburst with VLBI at multiple
frequencies. A first attempt has been carried out by Lobanov and Zensus (1996)
using data on 3C345 at 8 different frequencies ranging from 1.4 to 89 GHz.
Indeed a clear increase in peak frequency for two components was found which is
in contradiction with the basic shock models. Lobanov and Zensus
 also conclude that an
increase in the Doppler factor $\delta$ is the probable cause; however they
believe this is due to acceleration and/or a change in the angle to the line of
sight $\theta$. There is a hint for acceleration in the measured
angular velocities for these components (Zensus, Cohen and Unwin 1995), 
but the evidence for this is not convincing. 
If a change in the Doppler factor $\delta$ causes the increase in peak 
frequency (and peak flux density) in 3C345, it does not have to result 
in a change in the observed proper motion.
In figure \ref{deltabeta}, the influence of $\gamma$ and the
angle to the line of sight $\theta$ on both the Doppler factor $\delta$ and
the apparent proper motion $\beta_{app}$ is shown.
$\gamma$ varies for the solid lines while $\theta$ varies for the dotted
lines. The grey areas $A$ and $B$ indicate
the regions in parameter space for which deceleration and
acceleration respectively can cause the increase in Doppler boosting. 
If you follow a solid line in region $A$, $\delta$ increases 
while $\gamma$ decreases, the reverse is true in region $B$.
In region $A$ the change in apparent proper motion is
only small, while in region $B$, $\beta_{app}$  increases
significantly. Moreover in region $A$, the angle to the line of sight
$\theta$ is in the range $10^{\circ}-45^{\circ}$, while for region $B$, 
$\theta$ is in the range $0^{\circ}-10^{\circ}$. Hence, due to retardation
effects, an outburst undergoing an increase in $\delta$ due to acceleration
should be observed to have a much smaller timescale than an outburst
undergoing an increase in $\delta$ due to deceleration. The sources discussed 
in this paper are selected for their slow and well studied variability, and
are therefore most likely not at the smallest angles to the line of sight
 compared to other highly variable radio sources.
If this decelerated component model is correct, 
then no increase in peak frequency 
is expected for the rapidly variable sources with $\theta < 10^{\circ}$, 
unless $\gamma > 20$.
Note that a change in $\theta$, if it is combined with the right $\gamma$,
can also cause $\delta$ to increase without changing $\beta_{app}$ significantly.
However this is especially the case for small $\theta$, for which only a
small change in $\theta$ can already significantly increase $\delta$.

\section{Conclusions}
A qualitative model for GPS galaxies has been presented, 
in which components are expelled
from a nucleus at relativistic speeds, decelerate and possibly contribute to a
mini-lobe. At large angles to the line of sight,  the minilobes are dominant
because the Doppler factor of the young components is smaller than unity
along the line of sight direction. At small angles to the line of sight these
young components appear much brighter and become dominant. This model is
consistent with the symmetric angular morphology of the mini-lobes of GPS
galaxies and their low variability. It can explain the apparent lack of GPS
quasars at similar redshifts to the GPS galaxies. Furthermore it provides an
alternative explanation for the evolution of the spectral peak of flux density
outbursts in highly variable radio sources.
\section*{Acknowledgements}
We thank Margo Aller for providing us with data prior to publication.
We thank Walter Jaffe for useful comments on the manuscript.
{}

\end{document}